\documentclass[final]{IEEEphot}

\jvol{xx}
\jnum{xx}
\jmonth{MONTH}
\pubyear{YYYY}

\graphicspath{ {./images/} }
\usepackage{mathtools}
\usepackage{cite}
\usepackage{eqnarray,amsmath,amssymb,graphicx}
\usepackage{caption}
\usepackage{subcaption}

\DeclareMathOperator*{\argmax}{argmax}
\newcommand*{\argmaxl}{\argmax\limits}

\begin{document}

\title{Adaptive polarimetric image representation for contrast optimization of a polarized beacon through fog}

\author{S.~Panigrahi, J.~Fade, M.~Alouini}

\affil{Institut de Physique de Rennes, CNRS, Universit\'e de Rennes 1, Campus de Beaulieu, 35\,042 Rennes, France}  

\doiinfo{DOI: DOI.DOI/JPHOT.YYYY.XXXXXXX XXXX-XXXX/\$XX.XX \copyright YYYY}%

\maketitle


\begin{receivedinfo}%
Manuscript received MONTH DD, YYYY; revised MONTH DD, YYYY. First published MONTH DD, YYYY. Current version published MONTH DD, YYYY.
\end{receivedinfo}

\begin{abstract}
 We present a contrast-maximizing optimal linear representation of polarimetric images obtained from a snapshot polarimetric camera for enhanced vision of a polarized light source in obscured weather conditions (fog, haze, cloud) over long distances (above 1 km). We quantitatively compare the gain in contrast obtained by different linear representations of the experimental polarimetric images taken during rapidly varying foggy conditions. It is shown that the adaptive image representation that depends on the correlation in background noise fluctuations in the two polarimetric images provides an optimal contrast enhancement over all weather conditions as opposed to a simple difference image which underperforms during low visibility conditions. Finally, we derive the analytic expression of the gain in contrast obtained with this optimal representation and show that the experimental results are in agreement with the assumed correlated Gaussian noise model. 
\end{abstract}


\section{Introduction}

Polarimetric imaging produces multi-dimensional pixel data that is
either interpreted in terms of polarimetric properties of the imaged
objects, or quite often, processed into a single image revealing
specific contrasts which may not appear on standard reflectance
images. In simplified polarimetric imaging systems, a pair of
monochromatic images are acquired along two orthogonal polarization
directions. The two-dimensional pixel data obtained can then be
presented by either color encoding or by a combination of the
two components so as to enhance contrast between objects in a scene
sharing different polarimetric properties. Such contrast-maximizing representation of the 
polarimetric information can prove helpful in underwater polarimetric
imaging \cite{boffetyinfluence2012} and imaging through turbid media
\cite{ramachandrantwo-dimensional1998} (like colloids
\cite{demosoptical1997}, tissues \cite{nanlinear2009} and fog
\cite{AO:Expt}). In most laboratory based imaging experiments
\cite{Horinaka:95,Emile:96}, the scene is static and the object of
interest is usually embedded in a uniform background, thus allowing
for processing over multiple frames acquired over a period of
time. However, in real-world scenarios with fast moving scene and/or
camera, it is often desirable to reach real-time imaging and
processing. This requires the identification and use of
computationally simple and optimal representations of the polarimetric
images that are adapted to the experimental scenarios at hand.

In this article, we address the specific issue of contrast enhancement
of an intentionally polarized beacon of light (or semaphore), imaged at a
long distance through obscured atmosphere with a
polarization-sensitive camera. Such situation is of great interest for
applications in transportation safety. In this context, we
demonstrate, both analytically and experimentally, that an optimal
processing of the polarimetric images allows such contrast
maximization under all experimental conditions encountered. The
optimal polarimetric representation derived differs from commonly used
polarimetric contrasts, but remains computationally compatible with
real-time processing at video rate, which is a stringent constraint in
the applicative context considered here.


This article is organized as follows: in Section \ref{Exp}, the
experimental setup is described, as well as the statistical measure
used to assess the contrast of the source in the image. Then, the numerical and theoretical
derivation of an optimal polarimetric representation is reported in
Section \ref{optrep}. The efficiency of this optimal representation to enhance contrast of a
polarized source through fog is then discussed on experimental data in
Section \ref{discus}, allowing us to confirm theoretical predictions
in real field conditions. Conclusion and perspectives of this article
are finally given in Section \ref{conc}.

\section{Long distance polarimetric imaging experiment through fog}\label{Exp}

The long distance polarimetric imaging experiment described in the
following has been set up in the vicinity of the campus of
University of Rennes 1 to gather experimental data on real atmospheric
conditions. The imaging experiment covers a kilometric distance which
is the typical range of distance one aims at for transportation safety
applications like air and sea transport. A thorough description of this experimental facility and
of the snapshot polarimetric imager designed is reported in
\cite{AO:Expt}, along with a detailed depiction of the experiment
control and calibration procedure.

\subsection{Experimental setup}

    \begin{figure}
        \centering
        \begin{subfigure}{.5\textwidth}
        \centering
        \includegraphics[width=.99\linewidth]{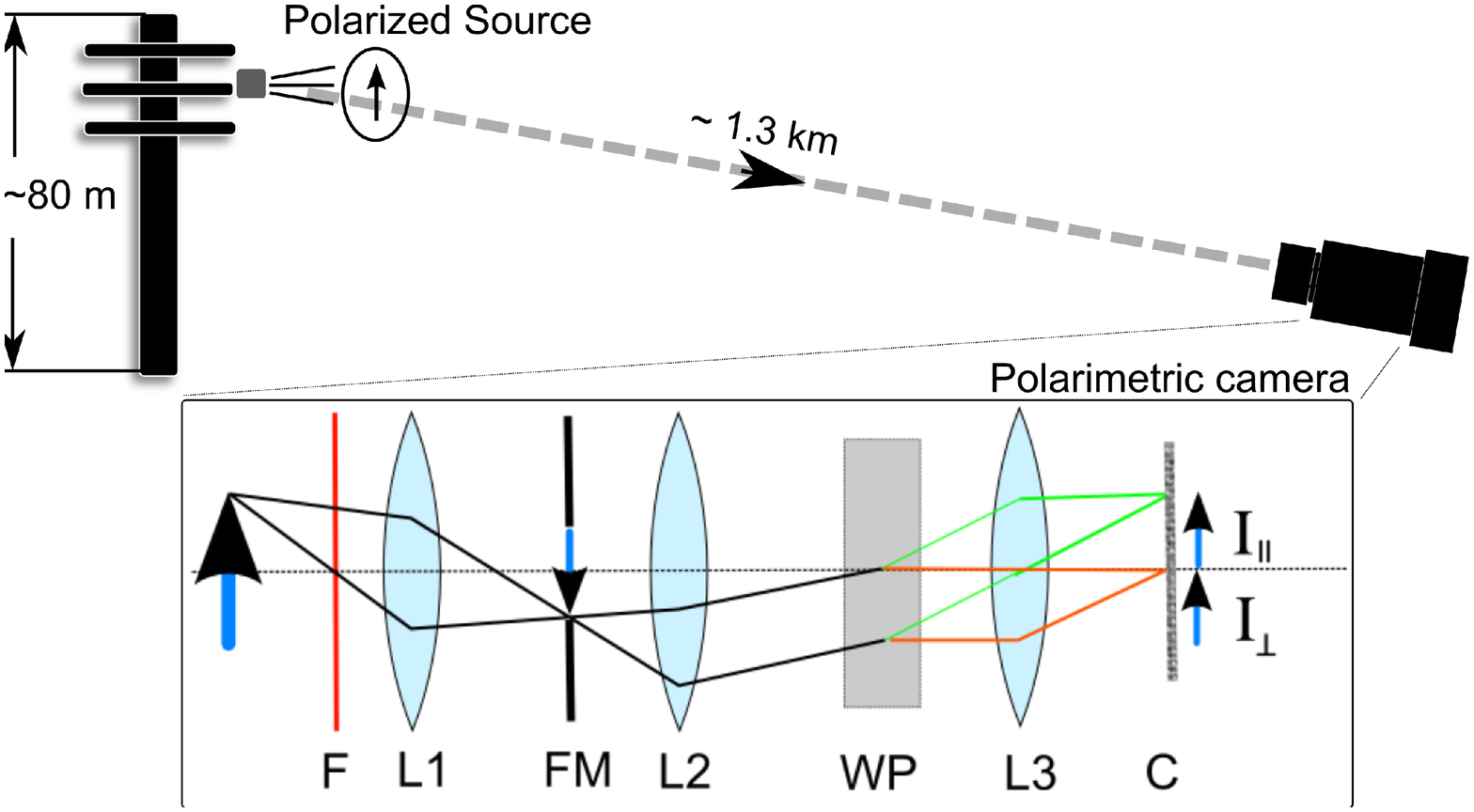}
        \caption{}
        \label{exptSetup}
        \end{subfigure}%
        \begin{subfigure}{.5\textwidth}
        \centering
        \includegraphics[width=.99\linewidth]{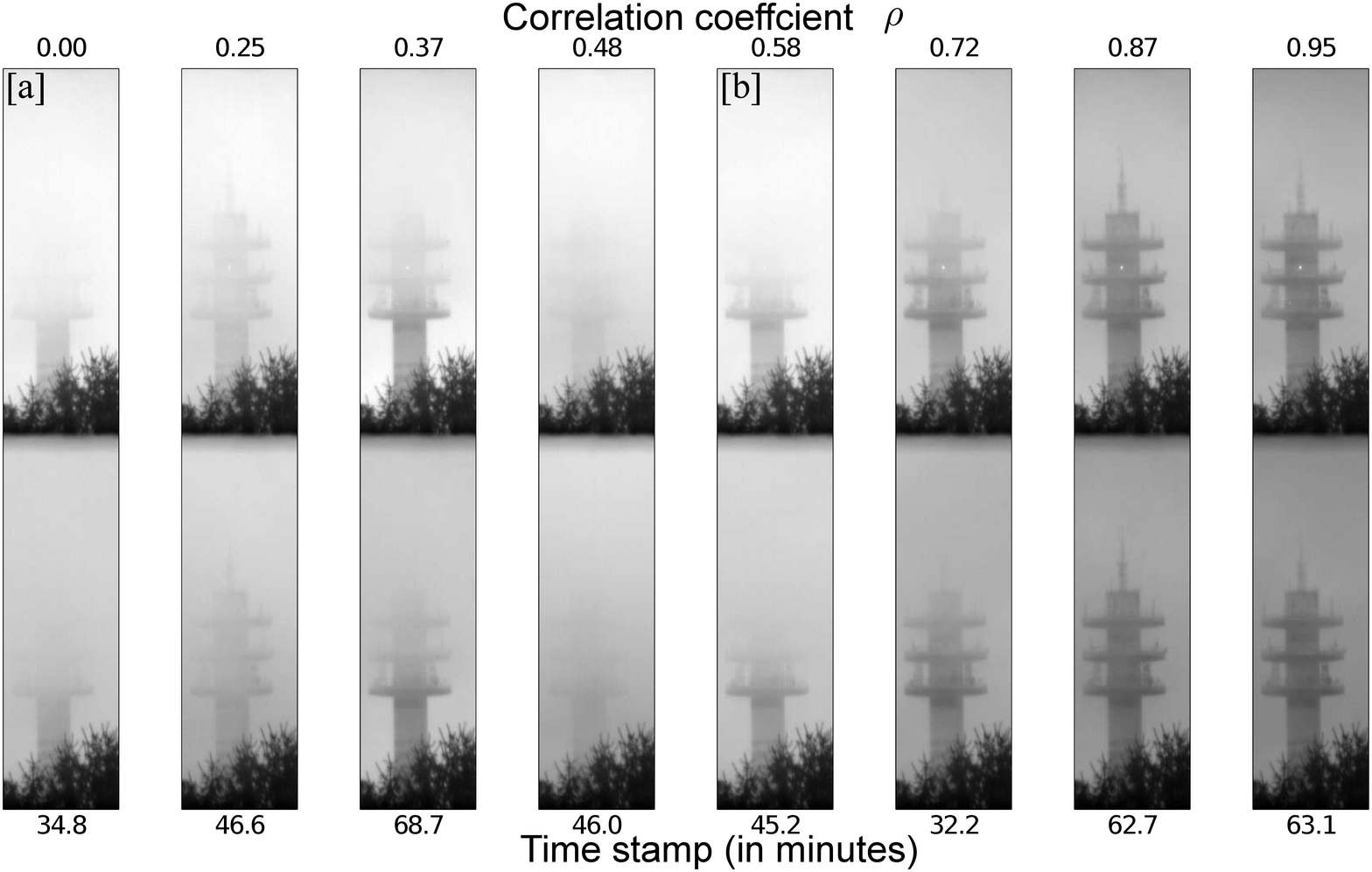}
        \caption{}
        \label{fig:raw}
        \end{subfigure}
        \caption{(a) The schematic shows the long range imaging setup.The polarimetric camera consists of the arrangement of lenses L1 and L2 after a monochromatic filter F (at 612 nm,
              FWHM = 12 nm). The image is partly masked by a slit (FM) and
              passed through a Wollaston prism (WP) to produce two images
              $I_\parallel$ and $I_\perp$ on a single camera (C) using
              lens L3. (b) A subset of the raw images from the polarimetric
              camera, showing the tower and the source with varying
              conditions of fog density and visibility. The 8 images are
              sorted in increasing order of background correlation ($\rho$)
              estimated over a small ROI surrounding the source pixel. The
              time stamp of acquisition is indicated below each
              image.}
    \end{figure}

    The experiment basically consists of a highly linearly polarized
    source of incoherent light, placed on a telecommunication tower
    (about 80~m in height and 1.3~km far from the detection site at the laboratory), and a Wollaston-prism based polarimetric
    camera for imaging. The experimental setup reported in
    \cite{AO:Expt} is improved here using a higher dynamics,
    low noise camera (Andor NEO sCMOS 5.5 Mpixels, 16-bit, 0.015
    e-/pixel/s dark noise at -30$^\circ$C sensor temperature) which is more
    suitable for this experiment. Such high-dynamics detector enables
    finer sampling of intensity levels and noise statistics in the
    acquired images. The entire imaging system has been thoroughly
    re-calibrated with this new camera.
    As mentioned before and shown in the schematic in Fig.~\ref{exptSetup}, the source is
    imaged from a distance of about 1.3 km, and the snapshot
    polarimetric imager enables the simultaneous acquisition of two
    images on the camera (namely, $I_\parallel$ and $I_\perp$)
    corresponding to the two orthogonal polarization directions, with
    $I_\parallel$ aligned with the direction of polarization of the
    source. 
    For the purpose of illustration in this article, we pick a dataset of images taken during an experiment
    conducted on 24-01-2014 between 1:00 p.m. and 2:20 p.m. (at a time
    interval of 1 frame/10 seconds).  During this 80 minutes 
    acquisition, successive passing fog layers obscured the source
    intermittently, causing the visibility of the source to evolve
    rapidly and significantly.  An example of 8 frames acquired by the
    polarimetric camera can be observed in Fig.~\ref{fig:raw}.

    \subsection{Polarimetric contrast image} \label{polConImg}

    Using the image registration method described in \cite{AO:Expt}, the two images can be extracted to form a set
    of two-dimensional pixels, such that the $i^{th}$ pixel $X^P_i=[ x_{\parallel,i}, x_{\perp,i} ]^T$
    is a part of the polarimetric image $I_P=\{X^P_i\}_{i\in[1,N]} = [I_\parallel, I_\perp]^T$. In practice, this two-dimensional data is processed to provide the
    end-user with a final contrast image, or to feed a higher-level
    image processing algorithm (detection and tracking, segmentation,
    etc.). For that purpose, the recorded two-dimensional data can be
    represented by a linear combination of both images $I_\parallel$
    and $I_\perp$, i.e., as a projection of the individual vectors
    $X^P_i$ over a row vector $W=[u,\,v]$. Thus, a linear representation
    denoted generically by $\gamma$ can be written as \begin{equation}\label{defrep}
      \gamma = W I_P= u\, I_\parallel + v\, I_\perp .\end{equation}

    In each such representation, the source will have different
    contrast and with different overall scaling depending on the
    values of $u$ and $v$. As a result, we resort to a
    contrast-to-noise ratio (CNR) in order to fairly compare the
    contrast of the source in each representation. Using a local
    region of interest (ROI) of size 21 $\times$ 21 pixels around the
    source, we identify two sets of pixels (shown as colored squares
    in Fig.~\ref{fig:visual} and described in the caption) denoted by
    $\mathcal{B}$ (\textit{background}) and $\mathcal{S}$
    (\textit{source}), and we define the CNR of the source in a
    general representation $\gamma$ as
    \begin{equation} \label{eq:CNR} \mathcal{C}(\gamma) = \left|
        \frac{\langle \gamma\rangle_{\mathcal{S}} - \langle
          \gamma\rangle_{\mathcal{B}}}
        {\widehat{\sigma}_\mathcal{B}(\gamma)} \right|,
    \end{equation}
    where, 
\begin{equation*}
  \langle \gamma \rangle _\chi=\sum_{i\in \chi} \frac{\gamma_i}{N_\chi}\text{\   and \   }\widehat{\sigma}_\chi^2(\gamma)= \frac{1}{N_\chi-1}
  \sum_{i\in \chi} \bigl(\gamma_i - \langle\gamma\rangle_{\chi}
  \bigr)^2
\end{equation*}
respectively stand for the empirical mean and variance
    over region $\chi$, with cardinality $N_\chi$.  
    This contrast measure returns the local contrast of a central pixel w.r.t. its immediate background and remains invariant under scaling of the gray levels in the image, i.e., for different values of $[u,v]$. In a general case, the intervening medium may be birefringent and thus the values of $u$ and $v$ can range between [-1,1]. However, for non-birefringent medium, where no rotation of polarization is observed, the weight ($u$) of $I_\parallel$ remains non-zero and thus can be scaled out so that the representation depends only on the weight ($v$) of $I_\perp$. Further, for a generic representation $\gamma$, we define a gain in contrast with respect to an intensity-summed image ($\gamma_\Sigma = I_\parallel + I_\perp$) which would be acquired with a standard camera. For brevity, we denote this gain as
    
\begin{equation}\label{eq:gainSym}
g^\Sigma=\frac{\mathcal{C}(\gamma)}{\mathcal{C}(\gamma_\Sigma)}.
\end{equation}

\section{Derivation of an optimal polarimetric representation}\label{optrep}

    Several combinations of the acquired
    polarimetric images can be envisaged for producing a final
    contrast image. In the context of this article, we aim at
    maximizing the CNR of a polarized light source over a
    background. This naturally raises the question of finding the
    optimal representation that provides the best contrast independent of the
    atmospheric situation, while remaining computationally efficient
    to match real-time requirements. Before deriving such optimal
    representation, let us first recall standard polarimetric
    representations which are commonly used in the literature for
    various applications in polarimetric imaging.

    \subsection{Standard polarimetric representations}

    From the two acquired polarimetric images $I_\parallel$ and
    $I_\perp$, most simple and standard representations are :

    \paragraph*{{Intensity-summed image ($\gamma_\Sigma= [1,\, 1]\,
        I_P$)}:} Such combination qualitatively provides the image
    that would be acquired with a standard, polarization-insensitive
    camera. Thus, other representations can be compared as a gain with respect to the intensity-summed image.
    \paragraph*{{Polarization filtered image ($\gamma_\parallel= [1,\,
        0]\, I_P$)}:} Another very simple approach is to use a
    polarization-filtered image, which can be obtained on our setup by
    retaining only the polarimetric image corresponding to the
    direction of polarization of the light source.
    \paragraph*{{Polarization-difference image ($\gamma_\Delta= [1,\,
        -1] \, I_P$)}:} Computing a difference image by subtracting the
    two polarimetric frames acquired is a very standard technique,
    widely used in polarimetric imaging \cite{tyotarget1996,Engheta-bio} for its
    efficiency in contrast enhancement. In the first experiments
    conducted with the imaging system described above, it was indeed
    noticed that the difference image performs generally better than
    the other standard representations \cite{AO:Expt}. Nevertheless,
    it was also observed that $\gamma_\Delta$ does not always provide
    the best possible contrast in the context of polarimetric imaging
    through fog considered in this article.
    \paragraph*{{Orthogonal States Contrast image ($OSC=\gamma_\Delta
        / \gamma_\Sigma$)}:} This polarimetric contrast is obtained by
    normalizing the difference image by the intensity-summed image. It
    is widely used and has proved efficient in active polarimetric
    imaging for its ability to provide an estimate of the degree of
    polarization of light scattered by an object (or an imaged
    scene)\cite{osc,osc2}. However, it was shown that such normalization tends to
    increase the noise in the final image due to possible low
    intensity values in the intensity-summed image \cite{AO:Expt}. For
    that reason, and due to the fact that the OSC cannot be written as
    a linear combination of $I_\parallel$ and $I_\perp$, this
    representation will not be considered in the remainder of this
    article.

    In the next subsections, we derive an optimal polarimetric representation, which in
    general differs from the most classical ones, whose expressions are
    summarized in Table \ref{tab}.

    \begin{figure}
        \centering
        \begin{subfigure}{.65\textwidth}
        \centering
        \includegraphics[width=.99\linewidth]{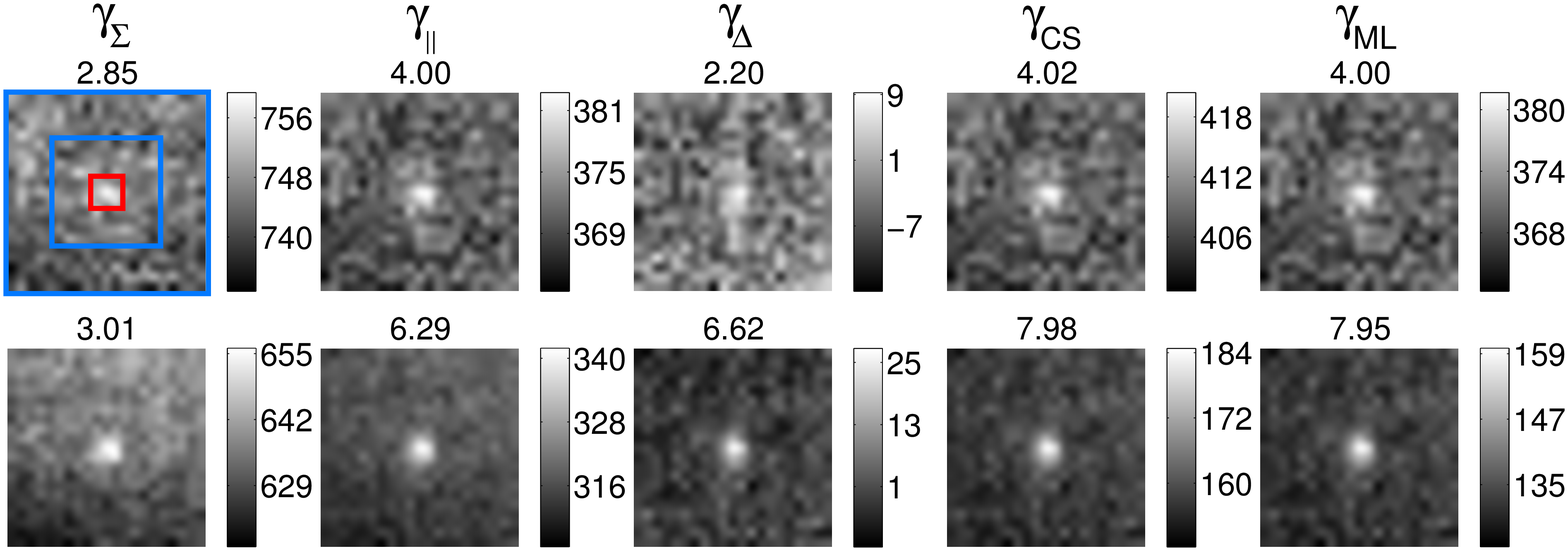}
        \caption{}
        \label{fig:visual}
        \end{subfigure}%
        \begin{subfigure}{.35\textwidth}
        \centering
        \includegraphics[width=.9\columnwidth]{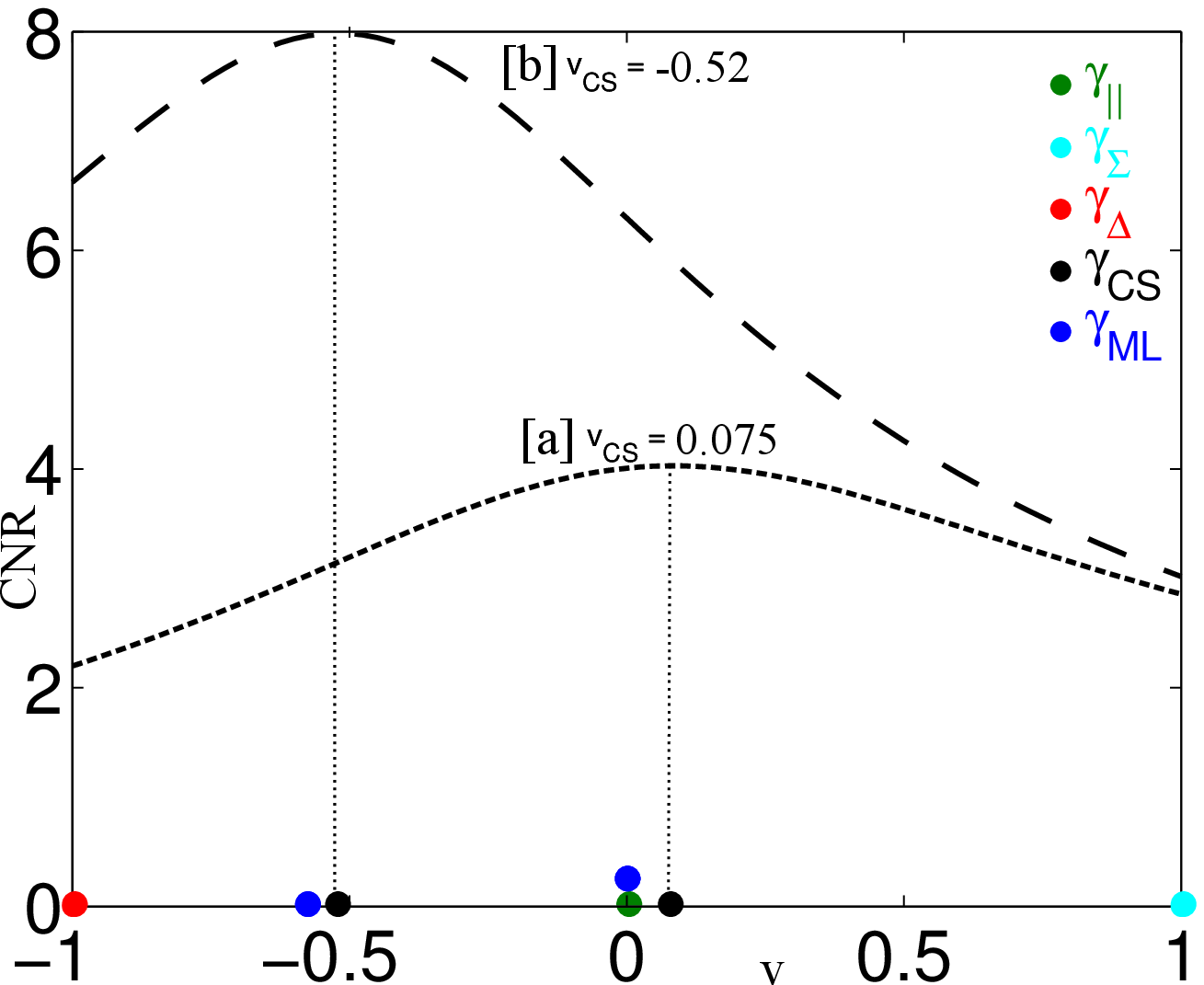}
        \caption{}
        \label{fig:CSsearch}
        \end{subfigure}
        \caption{(a) Comparison of the contrasts obtained for three representations of the polarimetric
          images for frames labeled as [a] and [b] in
          Fig.~\ref{fig:raw}. The source region, $\mathcal{S}$ is bounded by the 3 $\times$ 3 pixels red square and
          $\mathcal{B}$ is the background region between the two blue squares of sizes 11 $\times$ 11 pixels and 21 $\times$ 21 pixels. (b) CNR-maximizing 1D search over values of $v$ keeping $u=1$ for the two frames. }
    \end{figure}

 \subsection{Computational representation: Numerical maximization of CNR}
    
    It is possible to determine the optimal projection numerically
    for each acquired frame by a simple grid search over possible
    coefficient vectors $W$. As a result, the optimal representation,
    denoted $W _{CS}=[u_{CS},\, v_{CS}]$ in the following, is
    obtained by computationally solving $W_{CS}=\argmaxl_{W}\{ \mathcal{C}( W I_P)\},$
    for $u \in [0,1]$ and $v \in [-1,1]$ and obtaining an image $\gamma_{CS} = W_{CS} I_P$. 

    
    On the resulting processed images of two individual frames in
    Fig.~\ref{fig:visual}, on the ROI defined above, it can be checked that a significant
    contrast enhancement can be obtained over the difference image
    $\gamma_\Delta$ and the intensity-summed image $\gamma_\Sigma$,
    which indicates that the optimal representation $\gamma_{CS}$
    differs, in these cases, from both $\gamma_\Delta$ and
    $\gamma_\Sigma$. 
    It can be observed that for the frame
    labeled [a], the optimal representation is very close to a
    polarization-filtered image $\gamma_\parallel$. On the other hand,
    for the frame [b], the computational search leads to optimal
    weight of $v_{CS}$ corresponding to an intermediate
    situation between representations $\gamma_\Delta$ and
    $\gamma_\parallel$. 
   
    \subsection{Optimal representation: Theoretical maximization of CNR}    
    As observed in the previous subsection, the optimum linear combination of the polarimetric images for contrast enhancement may differ from the commonly used polarimetric representations, and may vary from frame to frame. Accordingly, identifying the physical parameter that influences the weights of the optimum linear combination would make it possible to implement an adaptive representation of polarimetric image that provides the best contrast for any weather condition. However, in such a long distance imaging setup, there is no a priori knowledge of the properties of the intervening medium, and thus we rely on the noise properties of the image. We hypothesize a correlated Gaussian noise model treating each pixel $X_i^P$ as a bivariate random variable having a mean of $\langle X^P \rangle_{\cal S} = [s + b/2 ,\ b/2]^T$ at the source location and $\langle X^P \rangle_{\cal B} = [b/2 ,b/2]^T$ outside the source location. Here, $s$ and $b$ denote the  mean intensities of the highly polarized source and the depolarized background, respectively. The second-order statistical properties of $X^P$ are modeled by the covariance matrix $\Gamma_i$ defined as
\begin{equation}\label{covar}
  \Gamma_i=\bigl\langle \bigl(X^P_i- \langle X^P_i\rangle \bigr)\bigl(X^P_i- \langle X^P_i\rangle \bigr)^T\bigr\rangle = \frac{\epsilon^2_i}{2}\begin{pmatrix}1 & \rho \\ \rho & 1\end{pmatrix},
\end{equation}
where $\epsilon_i$ stands for the standard deviation of the overall multiplicative optical noise, which is likely to be partially correlated in polarimetric channels especially in snapshot imaging. The correlation coefficient is denoted by $\rho$. With such statistical noise model, the theoretical expression of the CNR of a generic representation $\gamma$ has been derived in Appendix. The obtained expression can be easily and analytically maximized w.r.t. $\rho$, which indeed leads to the following linear representation $\gamma_{ML} = W_{ML} I_P$ with $W_{ML} = [1,-\rho]$. This provides a simple adaptive representation where the background noise correlation coefficient, $\rho$, which can be in practice estimated locally over the region $\mathcal{B}$ using the following empirical estimator

    \begin{equation} \label{eq:rho}
      \widehat{\rho} =  \sum_{ \mathclap{i\in \mathcal{B}} }
      \frac{
        (X_i^\parallel - \langle X^\parallel \rangle_\mathcal{B}) (X_i^\perp - \langle X^\perp \rangle_\mathcal{B})}
      {\widehat{\sigma_\mathcal{B}}(X^\parallel) \widehat{\sigma_\mathcal{B}}(X^\perp)},
    \end{equation}
    where ${\cal B}$ still denotes the background region comprised
    between the two blue squares of sizes 11 $\times$ 11 pixels and 21
    $\times$ 21 pixels depicted in Fig.~\ref{fig:visual}.
This representation is denoted as $\gamma_{ML}$ (for maximum likelihood) since, for the experimental conditions at hand (with highly polarized source and completely depolarized background), its form could be equivalently derived from a likelihood maximizing approach \cite{opex}. 
Using this representation on the same dataset, it can be seen in Fig.~\ref{fig:visual}.a that $\gamma_{CS}$ and $\gamma_{ML}$ consistently provide enhanced contrast compared to other simple polarimetric representations. As can be observed in Fig.~\ref{fig:visual}.b, these representations are almost equivalent, the small discrepancy between each other being due to numerical errors in the computation of the estimators and/or to possible deviation of the actual statistics from a Gaussian model. This result is verified over the entire dataset and further discussed in the next section.

    \section{Results and Discussion}    \label{discus}

\subsection{Experimental results}
In Fig.~\ref{fig:timeseriesGain}, we plot the time evolution of the gain in CNR defined in Eq.(\ref{eq:gainSym}) for each representation. For reference, the CNR ${\cal C}(\gamma_\Sigma)$ is shown as black-dotted line in the bottom of Fig.~\ref{fig:timeseriesGain}. The comparison confirms that the difference image $\gamma_\Delta$ (solid red lines) is not always the best representation and in many cases is outperformed by a simple `polarization-filtered' image $\gamma_\parallel$, i.e., the raw $I_\parallel$ image (solid, filled green). Furthermore, $\gamma_{CS}$ (solid black), presents the best contrast gain, and in general differs from both $\gamma_\Delta$ and $\gamma_\parallel$. It can be clearly observed that the gain in source contrast in the ML representation ($\gamma_{ML}$) closely follows the best possible gain obtained  with $\gamma_{CS}$, i.e., with a computational search over all possible linear combinations. As a result, this simple analytical representation behaves adaptively to present the best source contrast in the final image for all fog density conditions. 
These experimental results also quantify the advantage in using a polarimetric camera for long distance contrast-enhancement of a polarized beacon through fog, as the  CNR gain rises from 2-fold to a maximum of 12-fold compared to an intensity-summed image which is qualitatively similar to an image obtained from a standard intensity camera. It must be noticed at this level that the noise statistics of an intensity-summed image may differ in general from the ones obtained with a true intensity imager. As shown in Appendix, a fair comparison with a true intensity imager would imply a gain comprised between 2 and 6-fold, for the noise model considered.

   \begin{table}[!t]
    \centering
    \caption{Polarimetric representations and gains in CNR.}
    \label{tab}
    \begin{IEEEeqnarraybox}[\IEEEeqnarraystrutmode\IEEEeqnarraystrutsizeadd{2pt}{2pt}]{v/c/v/c/v/c/v/c/v}
      \IEEEeqnarrayrulerow\\
      &\mbox{{Representation}} && \mbox{{Symbol}} && W && g^\Sigma={\cal C}(\gamma)/{\cal C}(\gamma_\Sigma) \\
      \IEEEeqnarraydblrulerow\\
      \IEEEeqnarrayseprow[3pt]\\
      &\mbox{{Intensity-summed}}&& \gamma_\Sigma && [1,\,1] &&1\\
      \IEEEeqnarrayseprow[3pt]\\
      \IEEEeqnarrayrulerow\\
      \IEEEeqnarrayseprow[3pt]\\
      &\mbox{{Pol. filtered}}&& \gamma_\parallel && [1,\,0] && \sqrt{2(1+\rho)} \\
      \IEEEeqnarrayseprow[3pt]\\
      \IEEEeqnarrayrulerow\\
      \IEEEeqnarrayseprow[3pt]\\
      &\mbox{{Pol. difference}}&& \gamma_\Delta && [1,\,-1] && \sqrt{(1+\rho)/(1-\rho)}\\
      \IEEEeqnarrayseprow[3pt]\\
      \IEEEeqnarrayrulerow\\
      \IEEEeqnarrayseprow[3pt]\\
      &\mbox{{Computational}} && \gamma_{CS} && [1,\,v_{CS}] && - \\
      \IEEEeqnarrayseprow[3pt]\\
      \IEEEeqnarrayrulerow\\
      \IEEEeqnarrayseprow[3pt]\\
      &\mbox{{Max. Likelihood}}&& \gamma_{ML} && [1,\,-\rho] && \sqrt{2/(1-\rho)} \\
      \IEEEeqnarrayseprow[3pt]\\
      \IEEEeqnarrayrulerow
    \end{IEEEeqnarraybox}
    \end{table}\textbf{}
   \subsection{Influence of $\rho$ and theoretical gains in CNR}
\begin{figure}
        \centering
        \includegraphics[width=.5\columnwidth]{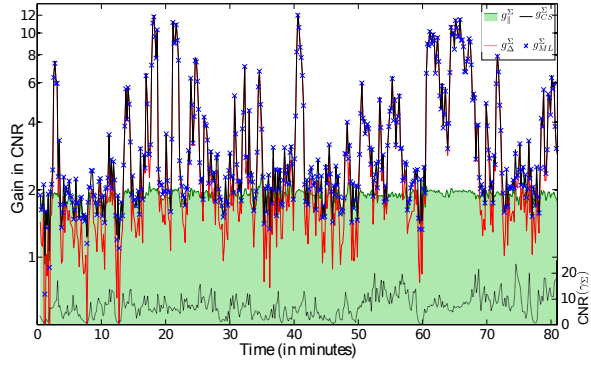}
        \caption{Gain in contrast (log scale) reached by the
          difference image $\gamma_\Delta$ (red solid lines) and
          $\gamma_{CS}$ representation (black solid line) w.r.t the intensity-summed image
          $\gamma_\Sigma$. The blue crosses show the gain obtained
          with $\gamma_{ML}$ representation, which can be seen to
          follow the maximum attainable contrast. The green filled
          curve shows the gain of the $\gamma_\parallel$ image with
          respect to the intensity-summed image. \label{fig:timeseriesGain} }
    \end{figure}

\begin{figure}
        \centering
        \begin{subfigure}{.38\textwidth}
        \centering
        \includegraphics[width=.95\linewidth]{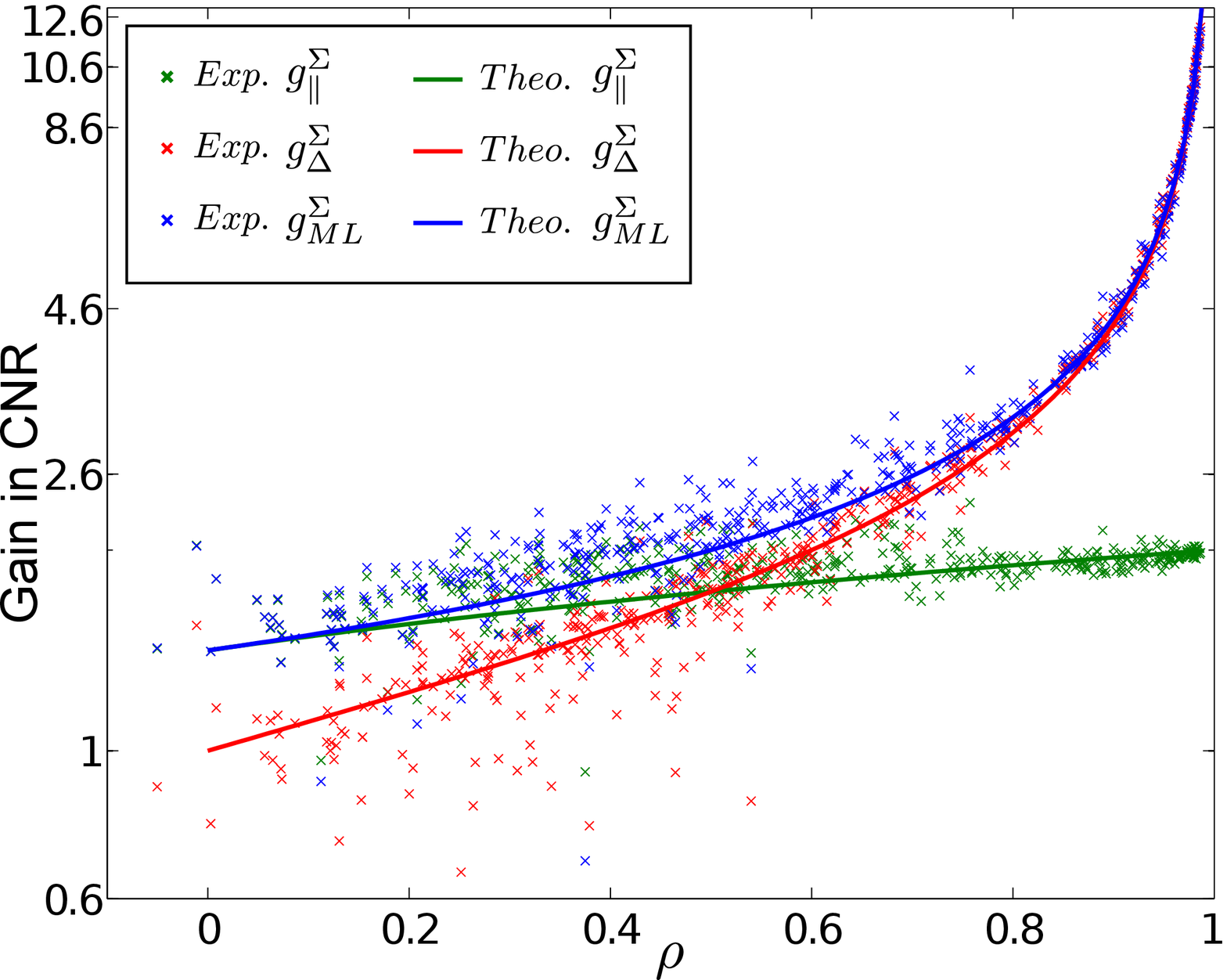}
        \caption{}
        \label{fig:corrGain}
        \end{subfigure}%
        \begin{subfigure}{.62\textwidth}
        \centering
        \includegraphics[width=.96\linewidth]{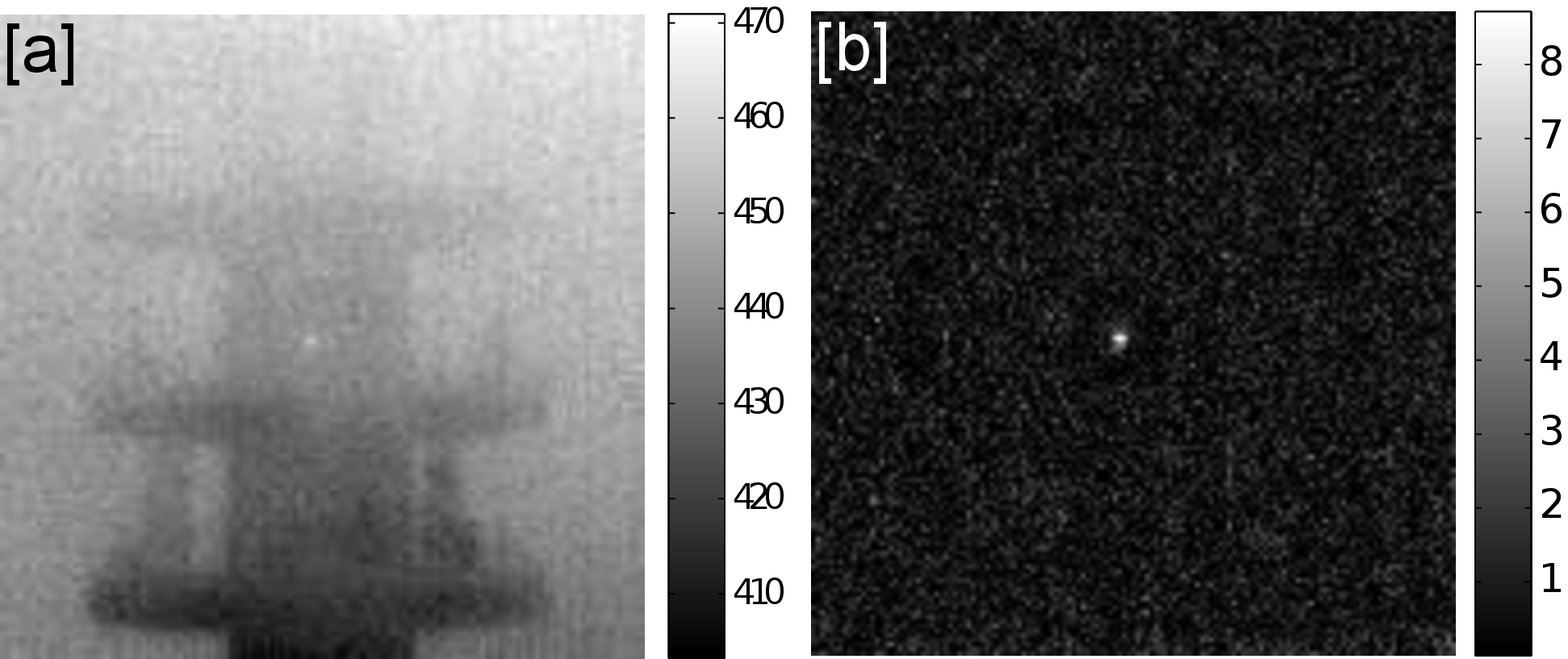}
        \caption{}
        \label{fig:CNRimage}
        \end{subfigure}
        \caption{(a) The experimentally obtained gain in CNR (scattered points) for each representation is plotted in log scale along-side the corresponding functional forms (solid lines) of the gain listed in Table~\ref{tab}. (b) Comparison of the intensity-summed image (left, [a]) with a processed CNR map of $\gamma_{ML}$ (right, [b]) that can be provided as a final contrast enhanced image to the end-user or to a higher-level image processing unit.\label{fig:Results} }
    \end{figure}

As stated in the previous section, the derivation of the ML representation allowed  us to identify the background correlation, $\rho$, as a crucial factor in determining the optimal contrast linear representation. In the above framework it is straightforward to calculate the theoretical CNR for each representation and thus compute the functional dependence of the gain in contrast with the correlation parameter (see Appendix). The theoretical forms are tabulated in Table ~\ref{tab} and plotted as solid lines in Fig.~\ref{fig:corrGain}. Furthermore, the experimentally generated contrast gains are plotted alongside (scattered symbols) as a function of locally estimated background correlation. 
    The plot shows that the CNR gain for each representation depends on
    $\rho$ in an orderly fashion which was not obvious in the `noisy'
    time-series data in Fig.~\ref{fig:timeseriesGain}.  It is
    interesting to notice that the difference image $\gamma_\Delta$
    gives the best CNR with high values of $\rho$, but is outperformed
    by the polarization-filtered image $\gamma_\parallel$ as the value
    of $\rho$ falls bellow 0.5. The performance of $\gamma_\parallel$
    remains linear with $\rho$, with a maximum gain of 2 for high values of
    $\rho$, while that of $\gamma_\Delta$ rises steeply when $\rho \to
    1$. Again, the optimality of the ML representation is clearly
    seen, as it corresponds to the best contrast representation for all
    values of $\rho$. \\
    The $\rho$-dependant performance of the studied representations (particularly $\gamma_\Delta$ and $\gamma_{ML}$) can be interpreted by noting that, in the present context,
    $\rho$ is a measure of the visibility of background structure in the
    local scene or local non-uniformity in reflectance. This can also be
    checked on the frames shown in Fig.~\ref{fig:raw} that are sorted
    in increasing order of $\rho$.  As a result, the main benefit of
    $\gamma_\Delta$ relies in its ability to suppress highly
    structured (and thus correlated) background in the final image. This property is retained by the $\gamma_{ML}$, which identifies with $\gamma_\Delta$ for $\rho \to 1$. On
    the other hand, with uncorrelated background ($\rho=0$), the
    perpendicular image $I_\perp$ does not bring any further
    information, making $\gamma_\parallel$ optimal during very low
    visibility conditions. The experiment also quantifies the advantage in using a polarimetric camera as we observe a maximum CNR gain of 12-fold compared to the intensity-summed image which is qualitatively similar to an image obtained from standard intensity camera.

   \subsection{Implementation of optimal representations}
Finally, we briefly discuss the implementation and some generic issues associated with computing such contrast-maximizing representations for each frame, when the locations and number of sources are unknown. In this case, each frame can be processed to obtain an intermediate contrast image that isolates the polarized source from the rest of the image as shown in Fig.~\ref{fig:CNRimage}. To obtain the CNR image, the calculation of CNR over a sliding window can be replaced with a convolution approach which performs quickly by using Fast Fourier Transforms. For computing $\gamma_{CS}$, multiple CNR images must be generated and the maximum value at each pixel must be chosen to form the final image. In contrast, for the ML representation one would need fewer Fourier transforms to calculate the local correlation coefficients, thus providing $\gamma_{ML}$ a noticeable time advantage over $\gamma_{CS}$. However, with fast computers or FPGA based embedded system, both techniques should remain within real-time requirement. Another parameter of importance is the size and shape of ROI, which should also be varied to maximize the CNR over multiple scales, especially when the spread of the source is unknown. The ML method remains specifically suitable in this case as only a 1D maximization over scale is required as opposed to a search over both scale and linear weighting of polarimetric images.

\textfloatsep 1.7\baselineskip

\section{Conclusion}\label{conc}

In this article, we first showed experimentally that when performing real field polarimetric imaging, two polarimetric channels acquired along two orthogonal polarization directions can have intensity fluctuations that are significantly correlated. Moreover, experiments reveal that the contrast of a polarized light source under any visibility condition can be maximized using a linear combination of the two acquired polarimetric images, which differs in general from the standard polarimetric representations used in literature. 
Under a correlated Gaussian noise hypothesis, we also demonstrated that the optimal representation is simply related to the noise correlation coefficient, which is also observed experimentally. As a result, such computationally-efficient representation can replace a numerical search of the optimal weighting coefficients, and could thus be easily implemented in real-time applications as a pre-processing task for automated detection/localization on wide field images. Lastly, the results presented here could be easily generalized to any case of partially polarized source and background with finite detector noise, which could be of interest for underwater imaging or imaging in biological tissues.


\bigskip

\appendix
\medskip

\renewcommand{\theequation}{\thesection-\arabic{equation}} 

As noted above, the polarimetric pixel can be written as a 2-D random
vector $X^P = [x^\parallel,x^\perp]^T$, where the dependency in pixel
location $i$ is omitted in the appendix for the sake of clarity. Assuming a correlated Gaussian noise model and keeping the same notations as in Section~\ref{polConImg}, we derive the expression for CNR for a generic polarimetric representation $\gamma$. Its mean value at a given location in the image is directly given by $\langle \gamma
\rangle=W\langle X^P \rangle$, and its variance reads
$\sigma^2(\gamma)=W \Gamma W^T$. From the definition of the CNR in
Eq.(\ref{eq:CNR}), a straightforward calculation yields
\begin{equation*} {\cal
    C}(\gamma)=\frac{s}{\epsilon}\sqrt{\frac{2u^2}{u^2+2\rho u v +v^2}}.
\end{equation*}
As a result, the gain in CNR with respect to the intensity-summed representation $\gamma_\Sigma$ reads
\begin{equation*} g^\Sigma=\frac{ {\cal C}(\gamma)}{{\cal
      C}(\gamma_\Sigma)}=\sqrt{\frac{2u^2(1+\rho)}{u^2+2\rho u v
      +v^2}},
\end{equation*}
from which the gain expressions of Table \ref{tab} are easily derived.

If one now considers a true intensity imager, and given the notations
above, the intensity level $X^I$ recorded at a given pixel would have
a mean value of $\langle X^I \rangle_{\cal S} =s+b$ at the source
location, and $\langle X^I \rangle_{\cal B} =b$ in the background
region, with variance $\sigma^2(X^I)=\epsilon^2$. The CNR is thus
${\cal C}(X^I)=s/\epsilon$, and as a result,
\begin{equation}
  g^I =\frac{{\cal C}(\gamma)}{{\cal C}(X^I)}= \frac{{\cal C}(\gamma)}{{\cal C}(\gamma_\Sigma)}\frac{{\cal C}(\gamma_\Sigma)}{{\cal C}(X^I)}= \frac{g^\Sigma}{\sqrt{1+\rho}}.
\end{equation} This shows that the gain
with respect to a true intensity imager is equivalent to the gain with
respect to the intensity-summed image only when the correlation parameter
$\rho$ tends to $0$, and is lower by a factor of 2 when $\rho$ approaches unity.
\textbf{}
\bigskip

\section*{Acknowledgments}
This work has been partly funded by the CEFIPRA (project N$^\circ$
4604-4). The authors would like to thank Pr. Hema Ramachandran for
fruitful discussions. The authors acknowledge the TDF company and
Rennes M\'{e}tropole for supporting this work, and L. Frein, C. Hamel,
A. Carr\'{e} and S. Bouhier for technical assistance.

\medskip





\end{document}